\documentstyle[epsf]{article}
\renewcommand{\cite}[1]{\ref{#1}}
\newcommand{\half}{\frac{1}{2}}
\begin{document}
\begin{flushleft}
{\bf A MECHANISM FOR A SMALL BUT NONZERO COSMOLOGICAL CONSTANT}\vspace{.9cm}\\
Yasunori FUJII\\
{\it Nihon Fukushi University, Handa, 470-32 Japan}, and \\
%and\\
{\it Institute for Cosmic Ray Research,
University of Tokyo, Tanashi, Tokyo, 188 Japan} \vspace{.5cm}\\
\end{flushleft}

I start with assuming that we do have a cosmological constant which is nearly as large as the critical density [\cite{os}];
\[
\Omega_{\Lambda}\equiv \frac{\Lambda}{3H_{0}^{2}}\sim 1.
\]
How can it be so in the context of unified theories?  This is the cosmological constant problem, which is in fact a 2-fold riddle\vspace{1em}:

$\bullet$ In almost any of the unification theories, we have a nonzero cosmological constant.  Unfortunately the observed value of $\Lambda$ is about 120 orders smaller than what we expect naturally from a theoretical point of view.  This is the first part of the 2-fold riddle.
If the above result on the size of $\Lambda$ were only an upper bound, it might be sufficient to invent a theory in such a way that the cosmological constant goes away entirely at least today.  An example is the simplest version of the decaying cosmological constant scenario [\cite{yf1}].

$\bullet$ Suppose, however, $\Lambda$ is in fact nonzero finite.  
Then we would face a tougher problem.  We have so small a number that we find it almost impossible to keep it undisturbed by any perturbation no matter how small.  This is the second part of the riddle\vspace{1em}.

The cosmological constant seems to be a very tiny but potentially extremely dangerous stumbling block on the road toward unification.
I will try to outline how I can evade it by proposing a theoretical model which I believe is not entirely unnatural.  
I support the view that there is a final theory that unifies everything at the Planck scale.  
But I do not like to base my argument on the exact details of unified theories which are still yet to be fully worked out.  
Instead I start with assuming an effective theory in 4 dimensions,  
expecting in particular that some scalar fields would play an important role.

In many of the models of unified theories we find scalar fields which couple to the ordinary matter as weakly as gravitation; they are different from Higgs fields having stronger interactions.  
An example is the so-called dilaton field.  
Another example is the scalar field representing a size of internal compactified space, a remnant of higher-dimensional spacetime.  
These scalar fields are characterized by the ``nonminimal'' coupling.  For the sake of illustration I give a simplified Lagrangian in 4-dimensions, as a reasonably good starting theory:
\begin{equation}
{\cal L}=\sqrt{-g}\left(  \half F(\phi) R  -\half (\partial\phi)^2
              -\Lambda +L_{\rm matter}  \right), \label{eq1}
\end{equation}
where $\phi$ is the scalar field.  $F(\phi)$ multiplied with $R$ is the nonminimal coupling, and $F(\phi)=\xi\phi^2$ is the simplest choice due originally to Brans and Dicke, with $\xi$ a constant related to BD's $\omega$ by $\xi\omega =1/4$ (also $\varphi_{BD}=\xi\phi^2 /2)$. 
I introduce $\Lambda$ at this level of the theory.  This imples that I focus upon the {\em primordial} cosmological constant, rather than the vacuum energies due to the cosmological phase transitions in the later epochs.

Also I use a unit system with
$c=\hbar = 8\pi G =1$. The present age of the Universe considered to be somewhere around 10 Gy is nearly $10^{60}$ in units of the Planck time $\sim 2.7\times 10^{-43}$sec.

Now we have to go through some of the complications concerning conformal transformations and conformal frames.  But I will skip all of them.  At the moment I simply apply a particular conformal transformation to remove the nonminimal coupling, so that the original Lagrangian is put into the new form:
\begin{equation}
{\cal L}=\sqrt{-g}\left(  \half R  
-\half (\partial\sigma)^2
               -\Lambda e^{-\sigma /\kappa } +L_{\rm matter}  \right). \label{eq2}
\end{equation}
There are a number of remarks relevant here.

Notice first the absence of the nonminimal coupling just as was designed.  Secondly, as a consequence, the {\it canonical} scalar field in this new conformal frame is $\sigma$ which is different from but is related to the original $\phi$ by
\[
\phi = e^{\sigma /4\kappa},   \label{eq3}
\]
with $\kappa$ a constant given by $\xi$.  Thirdly, the $\Lambda$-term is multiplied by 
$\phi^{-4}=e^{-\sigma /\kappa}$.  
This implies that the $\Lambda$-term now {\em acts as a potential of} $\sigma$.  
We may expect that $\sigma$ would fall off the exponential slope toward infinity.  
This would signal that any effect of $\Lambda$ will decrease with time.

That this is indeed the case can be shown by integrating the cosmological equations:
\begin{eqnarray}
&&3H^2 = \rho_{s}+\rho_{m}, \label{eq4a}\\
&&\ddot{\sigma}+3H\dot{\sigma}+\frac{dV}{d\sigma}=0, \label{eq4b}\\
&&\dot{\rho}_{m}+4H\rho_{m}=0, \label{eq4c}
\end{eqnarray}
where $\rho_{m}$ is the matter density, while $\rho_{s}$ is the density of the scalar field; 
\[
\rho_{s}=\half \dot{\sigma}^2 + V,
\]
with 
\[
V(\sigma)=\Lambda F(\phi)^{-2}.
\]

I here choose the nonminimal coupling $F(\phi)= 1+\xi\phi^2$ which is a little more complicated than BD's original suggestion; $V$ is now slightly flatter near the origin (see Fig. 1 of Ref. [\cite{yf1}]). 
This has an effect to bring about sufficient amount of the inflationary expansion of the primordial Universe.  

The numerical integration gives in fact solutions in which the scale factor $a(t)$ starts with an exponential growth but followed by the power-law behavior (see Fig. 2 of Ref. [\cite{yf1}]).  This implies that there is no trace of the truly constant $\Lambda$ in the asymptotic era.

Eq. (\ref{eq4a}) suggests that $\rho_{s}(t)$ may be interpreted as the effective cosmological constant $\Lambda_{\rm eff}$.  We also find that this $\Lambda_{\rm eff}$ behaves asymptotically like $\sim t^{-2}$, the same behavior as $\rho_{m}$, hence giving $\Lambda_{\rm eff}\sim 10^{-120}$ today ($t\sim 10^{60}$).  
This may provide an answer to the first part of the riddle, but certainly short of replying the second part.  The observations seem to tell us that $\rho_{s}$ must decrease like $t^{-2}$ as an overall behavior to ensure the size $\sim 10^{-120}$, but should deviate from the smooth fall-off, hopefully as a {\em leveling-off} behavior, which would {\em imitate} the constant $\Lambda$ at least locally.

But what kind of theory can do the job?  Without any clear clue, we decided to introduce another scalar field, called $\Phi$ [\cite{pl}].  
But I soon realized that merely introducing $\Phi$ is not enough; 
nothing spectacular will happen unless 
$\Phi$ is coupled to $\sigma$ in a nontrivial way.
Again without any useful guide, we tried several candidates based on a try-and-error basis.  Then we came across to an interction of the form
\begin{equation}
V(\sigma,\Phi)=e^{-\sigma /\kappa}\left[ \Lambda +
\half m^2\Phi^2 U(\sigma)\right],
\label{eq5}
\end{equation}
with
\begin{equation}
U(\sigma)=1+B\sin (\omega\sigma), \label{eq6}
\end{equation}
where $B$, $\omega$ and $m$ are constants.  This may look rather awkward.  If $B$ were zero, the second term of (\ref{eq5}) would be simply a mass term of the $\Phi$ field multiplied by $e^{-\sigma /\kappa}$ expected to come from the conformal transformation. 

With nonzero $B$, we find a parabolic slope in the $\Phi$ direction, whereas $V$ falls off exponentially in the $\sigma$ direction with the oscillation superimposed (see Fig. 4 of Ref. [\cite{pl}]).  
We may expect to roll down the potential slope toward $\sigma\rightarrow\infty$ probably with some smooth meandering behavior.
However, quite surprisingly, the solution shows something striking.  An example is shown in Fig. 1 [\cite{hwnt}].

The horizontal axis is $\log_{10}t$ in units of Planck time, so that we are now around 60.  I also gave initial conditions of the classical cosmology at $t_{1}$ which I tentativly chose $=10^{10}$.  
Also as an important rule of the game, I assume that all the constants in the theory as well as the initial values of the scalar fields are essentially of the order 1 in Planckian units.  

Then as we find, $\rho_{s}$ behaves as we wanted to see; an overall behavior $\sim t^{-2}$ and step-like leveling-offs.  Corresponding to each of them, we have a mini-inflation; a rapid but temporary rise of the scale factor.  I have two of them in this example.

I have adjusted parameters such that one of the mini-inflations shows up around the present epoch, as shwon in the zoomed-up view in Fig. 2.  We have $\Lambda_{\rm eff}$ which is {\em small today because our Universe is old, not because of a fine-tuning.}
I obtained the values $t_{0}$=12.1 Gy, $h= 0.81$ and $\Omega_{\Lambda}=0.67$, just as an illustration.  
  
I will discuss more about the characteristics of the solution.  We have two mini-inflations before the present epoch in this solution.  This number depends on the choice of parameters, however.  For somewhat different values of parameters, I have five of them, for example.
We also find that the whole behavior is a {\em nearly cyclic repetition} of the same pattern.  It is nearly periodical if we plot them against $ln t$ {\em rather than} $t$ itself.  This is a highly nonlinear effect, but I will try to give intuitive explanations.

Each pattern consists basically of two phases.  First a ``catapulting phase," in which both of the scalar fields are driven by the forces coming from the potential $V$.  $\sigma$ is pushed forward toward infinity while $\Phi$ toward the central valley at $\Phi =0$.  The increase of $\sigma$, however, makes the potential dwindle very quickly because of the factor $\exp(-\sigma /\kappa)$.  
For this reason and also due to the ``cosmological frictional forces" provided by $-3H\dot{\sigma}$ and $-3H\dot{\Phi}$, the scalar fields are soon decelerated.  

Now the $\sigma$ slowed down suffciently is trapped by the sinusoidal potential given by $\sin (\omega\sigma)$. In this way, the system enters eventually the ``dormant phase," during which both of the scalar fields come to almost complete stop, and $\rho_{s}$ stays constant, hence it is as if we had a truly constant $\Lambda$.  This is the key of our mechanism for a small but nonzero cosmological constant.

Interestingly, this leveling-off does not last forever.  The forces which once dwindled begin to build up again, bringing the system, rather likely, back again to the catapulting phase.  In this way, a set of the two phases would repeat itself, a process called ``recycling." 

In this connection I point out that the real origin of the recycling behavior lies in the dynamics of the system of two scalar fields coupled to eahc other; the cosmological environment plays only minor roles.

In fact in the isolated $\sigma$-$\Phi$ system, I find typical solutions as in Fig. 3, showing many repetitions indeed. 
If the same behavior takes place in the cosmological setting, then we would have mini-inflations, and consequently the effect of a small but nonzero $\Lambda$.

For some other parameters, however, I find solutions as in  Fig. 4.   Recycling ends prematually.  This is because the values of the scalar fields and their time-derivatives do not match sufficiently close to the previous values.  
If this premature ``derailing" occurs in the cosmological environment, the Universe would evolve just smoothly; no anomalous effect. These two behaviors are the typical ones though there are some other variants as well.  

The question is then how likely we get sufficiently long recycling.  Frankly speaking this is a hard question because I have too many parameters; in addition to the constants of the theory itself, like $B$ and $\omega$, I have at least four initial values of the two scalar fields. 
Nevertheless I have an impression that the chances of having a long recycling are rather high.

With the same model of the isolated $\sigma$-$\Phi$ model, I varied one of the initial values $\Phi_{1}$ with other values and constants held fixed.  In one such attempt, I changed $\Phi_{1}$ from 1.5 to 2.6 with equal spacing 0.05, obtaining 8 solutions (out of 23) showing sufficiently long recycling (see Fig. 5 of Ref. [\cite{hwnt}]).
This is a fair reflection of the general trend as far as I have tried.

To conclude we add a few more comments.  My story may have sounded somewhat complicated and messy.  But I point out that this mechanism has something in common with what is widely known as ``relaxation oscillation," which is happening in everyday life.  In playing violin, for example, one moves the bow rather slowly, still producing sounds of much higher frequencies.  Friction is obviously crucial.  There is no reason why something should not happen in Nature simply because it is complicated.  The same should be true also in the Universe.

As a generic feature of our solutions, I have several mini-inflations.  On this basis, I predict backward that the Universe may have experienced several $\Lambda$-dominated epochs.  This can be dangerous.
Suppose a calculation gives $\rho_{s}$ which is non-negligible compared with $\rho_{m}$ at the time of nucleosynthesis, for example.  It may jeopardize the success of the standard theory.
This has to be avoided.  This is in fact what I did when I selcted out the examples shown before.  In Fig. 1, for example, $\rho_{s}$ is kept below $\rho_{m}$ at $\log_{10}t\sim 45$, though $\rho_{s}$ may be dominant at other epochs.
This illustrates how I can constrain the theory by studying the past history of the Universe.  The same argument can be applied to the values of $H_{0}$ and $\Omega_{\Lambda}$ at the present epoch.  In any case my result at this moment is still away from the goal in terms of numerical fits.

I also admit that I have made several assumptions which I myself am not sure how to derive from more fundamental theories.  In this respect I emphasize that my approach is phenomenological.  I am asking what the fundamental theory should be like 
if the cosmological constant ceases to be a problem.
I hope with Weinberg [\cite{wnb}] that challenging the cosmological constant will open up a new breakthrough in our effort toward unification.
\bigskip
\begin{enumerate}
\item\label{os}See, for example, J.P. Ostriker and P.J. Steinhardt, {\it Cosmic Concordance}, and papers cited therein.  These authors suggest a set of representative values; $H_{0}=65$ km/sec/Mpc and $\Omega_{\Lambda}=0.65$.
\item\label{yf1}Y. Fujii and T. Nishioka, Phys. Rev. {\bf D42}, 361(1990), and papers cited therein.
\item\label{pl}Y. Fujii and T. Nishioka, Phys. Lett. {\bf B254}, 347(1991).
\item\label{hwnt}Y. Fujii, How natural is a small but nonzero cosmological constant? preprint, gr-qc/9508029; to be published in Particle Astrophysics.
\item\label{wnb}S. Weinberg, Rev. Mod. Phys. {\bf 61}, 1(1989).
\end{enumerate}
\newpage

\begin{figure}[h]
\hspace{-3cm}
\epsffile{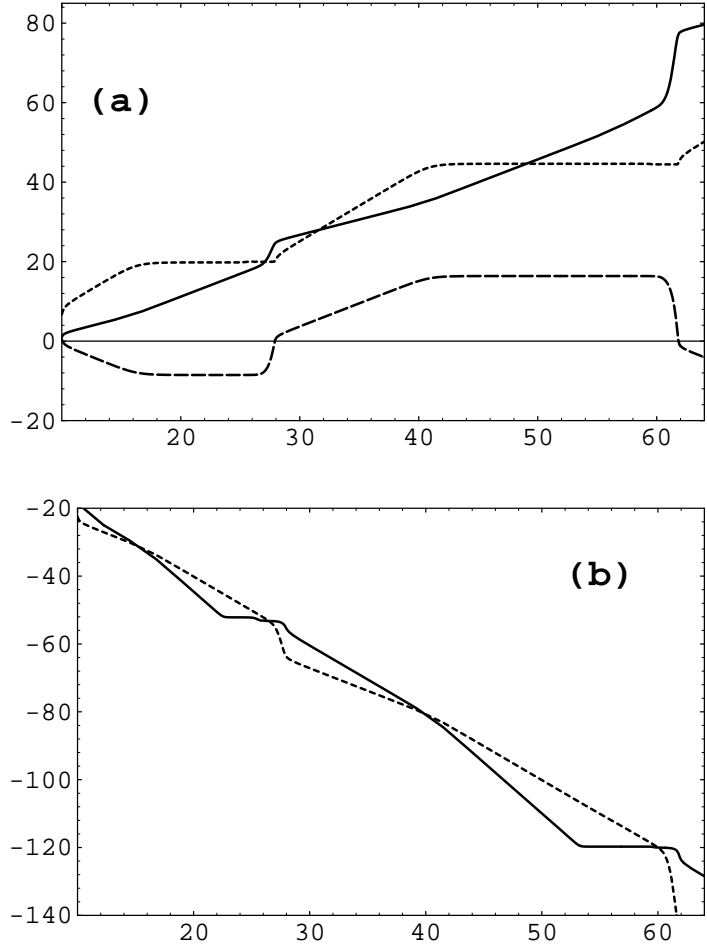}
\caption{An example of the solution of (3)-(5). (a) Upper plot: $b=\ln a$ (solid), $\sigma$ (dotted) and $2 \Phi$ (broken) are plotted against $\lambda \equiv\log_{10} t$. 
The present age of the Universe supposed to be (1.0\,-\,1.5)$\times 10^{10}$y corresponds to 60.0$\,$-$\,$60.2 of $\lambda$ in units of the Planck time.  
The parameters were chosen to be $\Lambda =1, \kappa =0.158, m=4.75, B=0.8, \omega =10$ in Planckian units.  
The initial values chosen conveniently at $t_{1}=10^{10}$ are $a=1, \sigma_{1}=6.75442, \dot{\sigma}_{1}=0, \Phi_{1}=0.212, \dot{\Phi}_{1}=0, \rho_{\rm r 1}=2.04\times 10^{-21}, \rho_{\rm nr 1}=4.46\times 10^{-44}$; the last two being adjusted to give the ``equal time" $\lambda_{\rm eq}\sim 55$.
The value of $\sigma_{1}$ corresponds to starting at a minimum of $\sin(\omega\sigma)$.  (b) Lower plot: $\rho_{\rm s}$ (solid), the total energy density of $\sigma$ and $\Phi$, and $\rho_{\rm m}$ (dotted), the matter energy density, against $\lambda$.  
}
\end{figure}

\begin{figure}[h]
\hspace{-3cm}
\epsffile{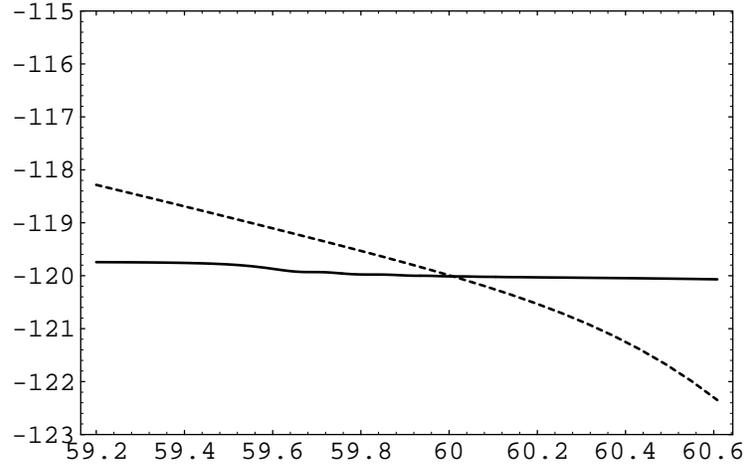}
\caption{The same plot as in Fig. 1(b) but in a magnified scale of $\lambda$ around the present time.
We find $\Omega_{\Lambda}=0.67$ and $H_{0}=81$km/sec/Mpc at $\lambda =60.15$ ($t=1.21\times 10^{10}$y).
}
\end{figure}

\begin{figure}[h]
\epsffile{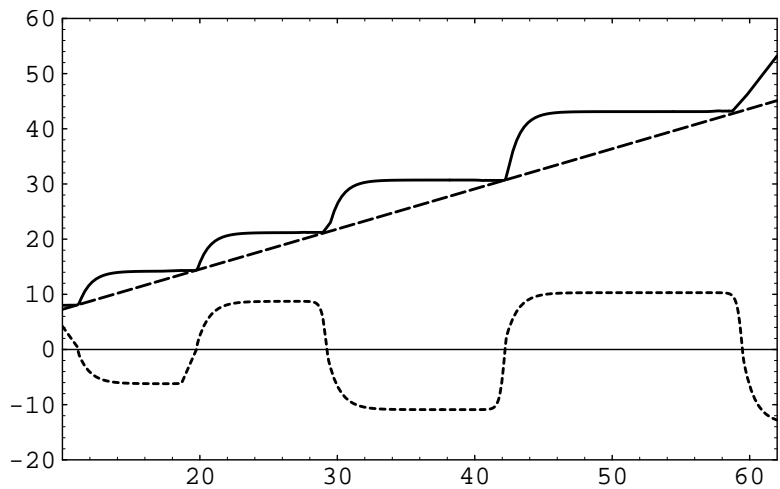}
\caption{An example of the solutions in the isolated $\sigma$-$\Phi$ system, showing long recycling.  
We choose $m=5.0$ with other constants as well as the symbols the same as in Fig. 1(a), except for $3b'-1$ replaced by 0.5, and the added broken line for $2\kappa\tau$ to be compared with $\sigma$.   
The initial values at $\lambda_{1}=10$ are $\sigma_{1}=8.0, \dot{\sigma}_{1}=0, \Phi_{1}=2.1, \dot{\Phi}_{1}=0.19$.
}

\end{figure}

\begin{figure}[h]
\epsffile{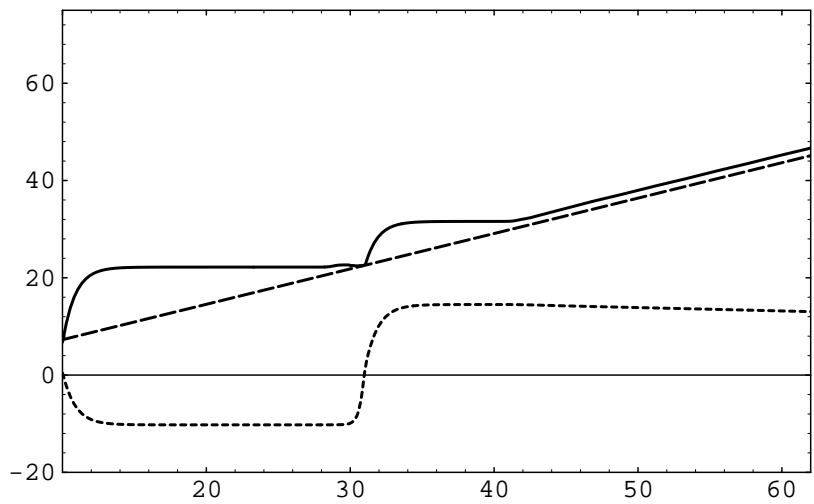}
\caption{An example of the solutions in the isolated $\sigma$-$\Phi$ system, in which recycling ends prematually at $\lambda\approx 41$. 
 The parameters and the initial values are the same as in Fig. 1
, except for $3b'-1$ replaced by 0.5.  
}
\end{figure}

\end{document}